\documentclass[twocolumn,showpacs,amsmath,amssymb,aps,prb,floatfix,nofootinbib,superscriptaddress]{revtex4}
\usepackage{graphicx,graphics,times}
\usepackage{bm}

\begin{document}
\title
{Controlling surface statistical properties using bias voltage:
Atomic force microscopy and stochastic analysis }

\author{P. Sangpour}
 \affiliation{Department of Physics,
Sharif University of Technology, P.O. Box 11365-9161, Tehran, Iran}

\author{G. R. Jafari}
 \affiliation{Department of Physics,
Sharif University of Technology, P.O. Box 11365-9161, Tehran, Iran}

\author{O. Akhavan}
  \affiliation{Department of Physics,
Sharif University of Technology, P.O. Box 11365-9161, Tehran, Iran}

\author{ A.Z. Moshfegh}
  \affiliation{Department of Physics,
Sharif University of Technology, P.O. Box 11365-9161, Tehran, Iran}

\author{ M. Reza Rahimi Tabar}
\affiliation{Department of Physics, Sharif University of Technology,
P.O. Box 11365-9161, Tehran, Iran}\affiliation{CNRS UMR 6529,
Observatoire de la C$\hat o$te d'Azur, BP 4229, 06304 Nice Cedex 4,
France}


\begin{abstract}

The effect of bias voltages on the statistical properties of rough
surfaces has been studied using atomic force microscopy technique
and its stochastic analysis.  We have characterized the complexity
of the height fluctuation of a rough surface by the stochastic
parameters such as roughness exponent, level crossing, and drift
and diffusion coefficients as a function of the applied bias
voltage. It is shown that these statistical as well as
microstructural parameters can also explain the macroscopic
property of a surface. Furthermore, the tip convolution effect on
the stochastic parameters has been examined.
 \pacs{ 05.10.Gg, 02.50.Fz, 68.37.-d}
\end{abstract}
\hspace{.3in}

 \maketitle
\section{Introduction}

As device dimensions continue to shrink into the deep submicron size
regime, there will be increasing attention for understanding the
thin-film growth mechanism and the kinetics of growing rough
surfaces in various deposition methods. To perform a quantitative
study on surfaces roughness, analytical and numerical treatments of
simple growth models propose, quite generally, the height
fluctuations have a self-similar character and their average
correlations exhibit a dynamic scaling form
\cite{Barabasi,Halpin,Krug,Meakin,Kardar,Masoudi}.  In these models,
roughness of a surface is a smooth function of the sample size and
growth time (or thickness) of films.  In addition, other statistical
quantities such as the average frequency of positive slope level
crossing, the probability density function (PDF), as well as drift
and diffusion coefficients provide further complete analysis on
roughness of a surface. Very recently, it has been shown that, by
using these statistical variables in the Langevin equation,
regeneration of rough surfaces with the same statistical properties
of a nanoscopic imaging is possible \cite{Jafari,Waechter}.

In practice, one of the effective ways to modify roughness of
surfaces is applying a negative bias voltage during deposition of
thin films \cite{Ohring}, while their sample size and thickness
are constant. In bias sputtering, electric fields near the
substrate are modified to vary the flux and energy of incident
charged species. This is achieved by applying either a negative DC
or RF bias to the substrate. Due to charge exchange processes in
the anode dark space, very few discharge ions strike the substrate
with full bias voltage. Rather a broad low energy distribution of
ions bombard the growing films.

Generally, bias sputtering modifies film properties such as
surface morphology, resistivity, stress, density, adhesion, and so
on through roughness improvement of the surface, elimination of
interfacial voids and subsurface porosity, creation of a finer and
more isotropic grain morphology, and the elimination of columnar
grains \cite{Ohring}.

In this work, the effect of bias voltage on the statistical
properties of a surface, i.e., the roughness exponent, the level
crossing, the probability density function, as well as the drift
and diffusion coefficients has been studied.  In this regard, we
have analyzed the surface of Co(3 nm)/NiO(30 nm)/Si(100) structure
(as a base structure in the magnetic multilayers, e.g., spin
valves operated using giant magnetoresistance (GMR) effect
\cite{Egelhoff,Dai}) fabricated by bias sputtering method at
different bias voltages. The behavior of statistical
characterizations obtained by nanostructural analysis have been
also compared with behavior of sheet resistance measurement of the
films deposited at the different bias voltages, as a macroscopic
analysis.

\section{Experimental}

The substrates used for this experiment were n-type Si(100) wafers
with resistivity of about 5-8 $\Omega$-cm and the dimension of
5$\times$11 mm$^2$ . After a standard RCA cleaning procedure and a
short time dip in a diluted HF solution, the wafers were loaded
into a vacuum chamber.  The chamber was evacuated to a base
pressure of about 4$\times$10$^{-7}$ Torr. To deposit nickel oxide
thin film, first high purity NiO powder was pressed and baked over
night at 1400 $^{\circ}$C in an atmospheric oven yielded a green
solid disk suitable for thermal evaporation.  Before each NiO
deposition, a pre-evaporation was done for about 5 minutes.  Then
a 30 nm thick NiO layer was deposited on the Si substrate with
applied power of about 350 watts resulted in a deposition rate of
0.03 nm/s at a pressure of 2$\times$10$^{-6}$ Torr.  After that,
without breaking the vacuum, a thin Co layer of 3 nm was deposited
on the NiO surface by using DC sputtering technique.  During the
deposition, a dynamic flow of ultrahigh purity Ar gas with
pressure of 70 mTorr was used for sputtering discharge. The
discharge power to grow Co layers was considered around 40 watts
that resulted in a deposition rate of about 0.01 nm/s.  The
thickness of the deposited films was measured by styles technique,
and controlled in-situ by a quartz crystal oscillator located near
the substrate. The distance between the target (50 mm in diameter)
and substrate was 70 mm. Before each deposition, a pre-sputtering
was also performed for about 10 minutes. The deposition of Co
layers was done at various negative bias voltages ranging from
zero to -80 V at the same sputtering conditions.  The schematic
details about the way of exerting the bias voltage to the Si
substrate can be found in \cite{AkhavanJPD}.

In order to analyze the deposited samples, we have used atomic force
microscopy (AFM) on contact mode to study the surface topography of
the Co layer. The surface topography of the films was investigated
using Park Scientific Instruments (model Autoprobe CP). The images
were collected in a constant force mode and digitized into $256
\times 256 $ pixels with scanning frequency of $0.6$ Hz. The
cantilever of 0.05 N $m^{-1}$ spring constant with a commercial
standard pyramidal Si$_3$N$_4$ tip with an aspect ratio of about 0.9
was used. A variety of scans, each with size $L$ were recorded at
random locations on the Co film surface. The electrical property of
the deposited films was examined by four-point probe sheet
resistance (R$_s$) measurement at room temperature.

\section{Statistical quantities}

\subsection {roughness exponents}
It is known that to derive a quantitative information of a surface
morphology one may consider a sample of size $L$ and define the
mean height of growing film $\overline{h}$ and its roughness $w$
by the following expressions \cite{Marsilli}:
\begin{equation}
\overline{h}(L,t,\lambda)=\frac{1}{L}\int_{-L/2}^{L/2}h(x,t,\lambda)dx
\end{equation}
and
\begin{equation}\label{w}
w(L,t,\lambda) =(\langle (h-\overline{h})^2\rangle)^{1/2}
\end{equation}
where $t$ is proportional to deposition time and
$\langle\cdots\rangle$ denotes an averaging over different
samples, respectively. Moreover, we have introduced $\lambda$ as
an external factor which can apply to control the surface
roughness of thin films.  In this work, $\lambda\equiv V/V_{opt}$
is defined where $V$ and $V_{opt}$ are the applied and the optimum
bias voltages, so that at $\lambda=1$ the surface shows its
optimal properties.  For simplicity, we assume that
$\overline{h}=0$, without losing the generality of the subject.
Starting from a flat interface (one of the possible initial
conditions), we conjecture that a scaling of space by factor $b$
and of time by factor $b^z$ ($z$ is the dynamical scaling
exponent), rescales the roughness $w$ by factor $b^{\chi}$ as
follows:
\begin{equation}\label{scaling}
w(bL,b^zt,\lambda)=b^{\chi(\lambda)}w(L,t,\lambda)
\end{equation}
which implies that
\begin{equation}
w(L,t,\lambda)=L^{\chi(\lambda)}f(t/L^z,\lambda).
\end{equation}
If for large $t$ and fixed $L$ $(t / L^z \rightarrow \infty)$ $w$
saturate, then $f(x,\lambda)\longrightarrow g(\lambda)$, as
$x\longrightarrow\infty$. However, for fixed large $L$ and
$t<<L^z$, one expects that correlations of the height fluctuations
are set up only within a distance $t^{1/z}$ and thus must be
independent of $L$. This implies that for $x << 1$, $f(x)\sim
x^{\beta}g'(\lambda)$ with $\beta=\chi / z$. Thus dynamic scaling
postulates that
\begin{eqnarray}
w(L,t,\lambda)=
\left\{%
\begin{array}{ll}
    t^{\beta(\lambda)}g(\lambda)\sim t^{\beta(\lambda)}, & \hbox{t$\ll L^z$;}\\
    L^{\chi(\lambda)}g'(\lambda)\sim L^{\chi(\lambda)}, & \hbox{t$\gg L^{z}$}. \\
\end{array}%
\right.
\end{eqnarray}
The roughness exponent $\chi$ and the dynamic exponent $z$
characterize the self-affine geometry of the surface and its
dynamics, respectively. The dependence of the roughness  $w$ to
the $\overline{h}$ or $t$ shows that $w$ has a fixed value for a
given time.

The common procedure to measure the roughness exponent
 of a rough surface is use of a surface structure function depending
on the length scale $\triangle x=r$ which is defined as:
\begin{eqnarray}\label{Structure}
S(r)=\langle|h(x+r)-h(x)|^2\rangle.
\end{eqnarray}
It is equivalent to the statistics of height-height correlation
function $C(r)$ for stationary surfaces, i.e. $S(r)=2w^2(1-C(r))$.
The second order structure function $S(r)$, scales with $r$ as $
r^{\xi_2}$ where $\chi=\xi_2 /2$ [1].

\subsection{ The Markov nature of height fluctuations}

 We have examined whether the data of height fluctuations follow a Markov
chain and, if so, determine the Markov length scale $l_M$. As is
well-known, a given process with a degree of randomness or
stochasticity may have a finite or an infinite Markov length scale
\cite{Peinke04}. The Markov length scale is the minimum length
interval over which the data can be considered as a Markov
process. To determine the Markov length scale $l_M$, we note that
a complete characterization of the statistical properties of
random fluctuations of a quantity $h$ in terms of a parameter $x$
requires evaluation of the joint PDF, i.e.
$P_N(h_1,x_1;....;h_N,x_N)$, for any arbitrary $N$. If the process
is a Markov process (a process without memory), an important
simplification arises. For this type of process, $P_N$ can be
generated by a product of the conditional probabilities
$P(h_{i+1},x_{i+1}|h_i,x_i)$, for $i=1,...,N-1$. As a necessary
condition for being a Markov process, the Chapman-Kolmogorov
equation,
\begin{eqnarray}
  &&P(h_2,x_2|h_1,x_1)= \cr \nonumber\\
  &&\int \hbox{d} (h_i)\,
  P(h_2,x_2|h_i,x_i)\, P(h_i,
  x_i| h_1,x_1)
\end{eqnarray}
should hold for any value of $x_i$, in the interval $ x_2<x_i<x_1$
\cite{Risken}.

The simplest way to determine $l_M$ for stationary or homogeneous
data is the numerical calculation of the quantity,
$S=|P(h_2,x_2|h_1,x_1)-\int\hbox{d}
h_3P(h_2,x_2|h_3,x_3)\,P(h_3,x_3|h_1,x_1)|$, for given $h_1$ and
$h_2$, in terms of, for example, $x_3-x_1$ and considering the
possible errors in estimating $S$. Then, $l_M=x_3-x_1$ for that
value of $x_3-x_1$ such that, $S=0$.

 It is well-known that the Chapman-Kolmogorov
equation yields an evolution equation for the change of the
distribution function $P(h,x)$ across the scales $x$. The
Chapman-Kolmogorov equation formulated in differential form yields
a master equation, which can take the form of a Fokker-Planck
equation \cite{Risken}:
\begin{eqnarray}\label{Fokker}
  \frac {d}{d r} P(h,x)=
 [-\frac{\partial }{\partial h}
  D^{(1)}(h,x)
  +\frac{\partial^2 }{\partial h^2} D^{(2)}(h,x)]
  P(h,x).
\end{eqnarray}
The drift and diffusion coefficients $D^{(1)}(h, r)$, $D^{(2)}(h,
r)$ can be estimated directly from the data and the moments
$M^{(k)}$ of the conditional probability distributions:
\begin{eqnarray}\label{D(k)}
  && D^{(k)}(h,x) = \frac{1}{k!}
   {\rm lim}_{r \rightarrow 0}  M^{(k)} \cr \nonumber \\
  && M^{(k)} = \frac{1}{r}  \int dh'
  (h'-h)^k P(h',x+r|
  h,x).
\end{eqnarray}
The coefficients $D^{(k)}(h,x)$`s are known as Kramers-Moyal
coefficients.  According to Pawula`s theorem \cite{Risken}, the
Kramers-Moyal expansion stops after the second term, provided that
the fourth order coefficient $D^{(4)} (h,x)$ vanishes
\cite{Risken}.
 The forth order coefficients $D^{(4)}$
 in our analysis was found to be about $ {D^{(4)}} \simeq 10^{-4} {D^{(2)}}$. In this
approximation, we can ignore the coefficients $D^{(n)}$ for $n
\geq 3$.

Now, analogous to equation (\ref{Fokker}), we can write a
Fokker-Planck equation for the PDF of $h$ which is equivalent to
the following  Langevin equation (using the Ito interpretation)
\cite{Risken}:
\begin{equation}\label{Langevin}
  \frac{d}{d x}  h(x,\lambda)=D^{(1)}(h,x,\lambda) +
  \sqrt{D^{(2)}(h,x,\lambda)}f(x)
\end{equation}
where $f(x)$ is a random force, zero mean with gaussian
statistics, $\delta$-correlated in $x$, i.e. $\langle
f(x)f(x')\rangle=\delta(x-x')$. Furthermore, with this last
expression, it becomes clear that we are able to separate the
deterministic and the noisy components of the surface height
fluctuations in terms of the coefficients $D^{(1)}$ and $D^{(2)}$.

\subsection{The level crossing analysis}

We have utilized the level crossing analysis in the context of
surface growth processes, according to \cite{Level,Shahbazi}. In the
level crossing analysis, we are interested in determining the
average frequency (in spatial dimension) of observing of the
definite value for height function $h=\alpha $ in the thin films
grown at different bias voltages, $\nu _\alpha ^{+} (\lambda)$.
Then, the average number of visiting the height $h=\alpha$ with
positive slope in a sample with size $L$ will be $N_\alpha ^{+}
(\lambda)=\nu _\alpha ^{+} (\lambda) L$. It can be shown that the
$\nu _\alpha ^{+}$ can be written in terms of the joint PDF of $h$
and its gradient. Therefore, the quantity $\nu _\alpha ^{+}$ carry
the whole information of surface which lies in $P(h, {h}^{\prime})$,
where $h'=dh/dx$, from which we get the following result for the
frequency parameter $\nu_{\alpha}^{+}$ in terms of the joint
probability density function
\begin{equation}
\nu_{\alpha}^{+}
=\int_{0}^{\infty}p(\alpha,{h}^{\prime}){h}^{\prime}d{h}
^{\prime}.
\end{equation}
The quantity $N_{tot}^{+}$ which is defined as $
N_{tot}^{+}=\int_{-\infty }^{+\infty }\nu _\alpha ^{+}d\alpha $ will
measure the total number of crossing the surface with positive
slope.  So, the $N_{tot}^{+}$  and square area of growing surface
are in the same order.   Concerning this, it can be utilized as
another quantity to study further the roughness of a surface. It is
expected that in the stationary state the $N_{tot}^{+}$  depends on
bias voltages.

\section{Results and discussion}

To study the effect of the bias voltage on the surface statistical
characteristics, we have utilized AFM method for obtaining
microstructural data from the Co layer deposited at the different
bias voltages in the Co/NiO/Si(100) system. Figure 1 shows AFM
micrographs of the Co layer deposited at various negative bias
voltages of -20, -40, -60, and -80 V, as compared with the
unbiased samples.

\begin{figure}
\includegraphics[width=6cm,height=6cm,angle=0]{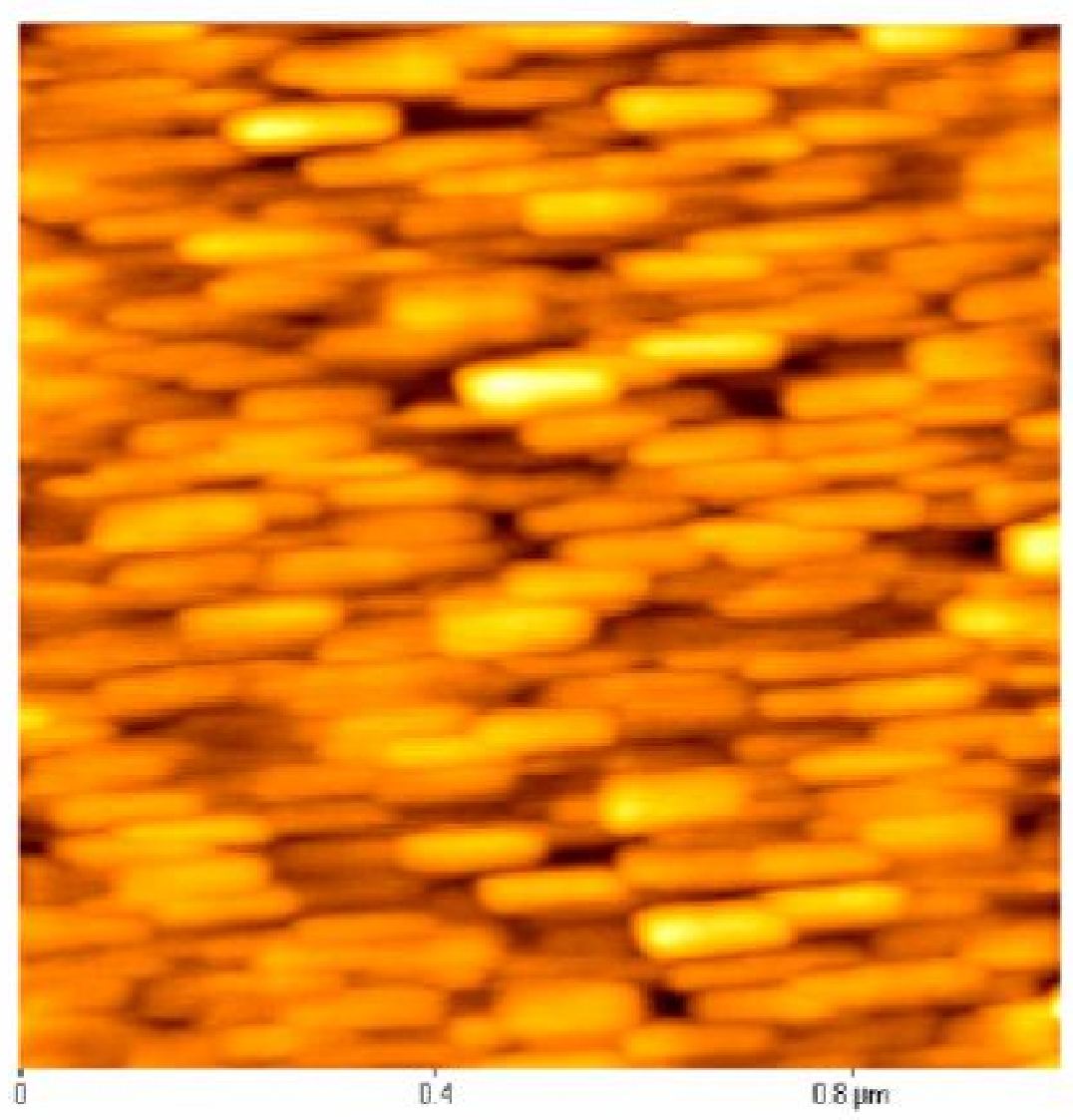}
\includegraphics[width=6cm,height=6cm,angle=0]{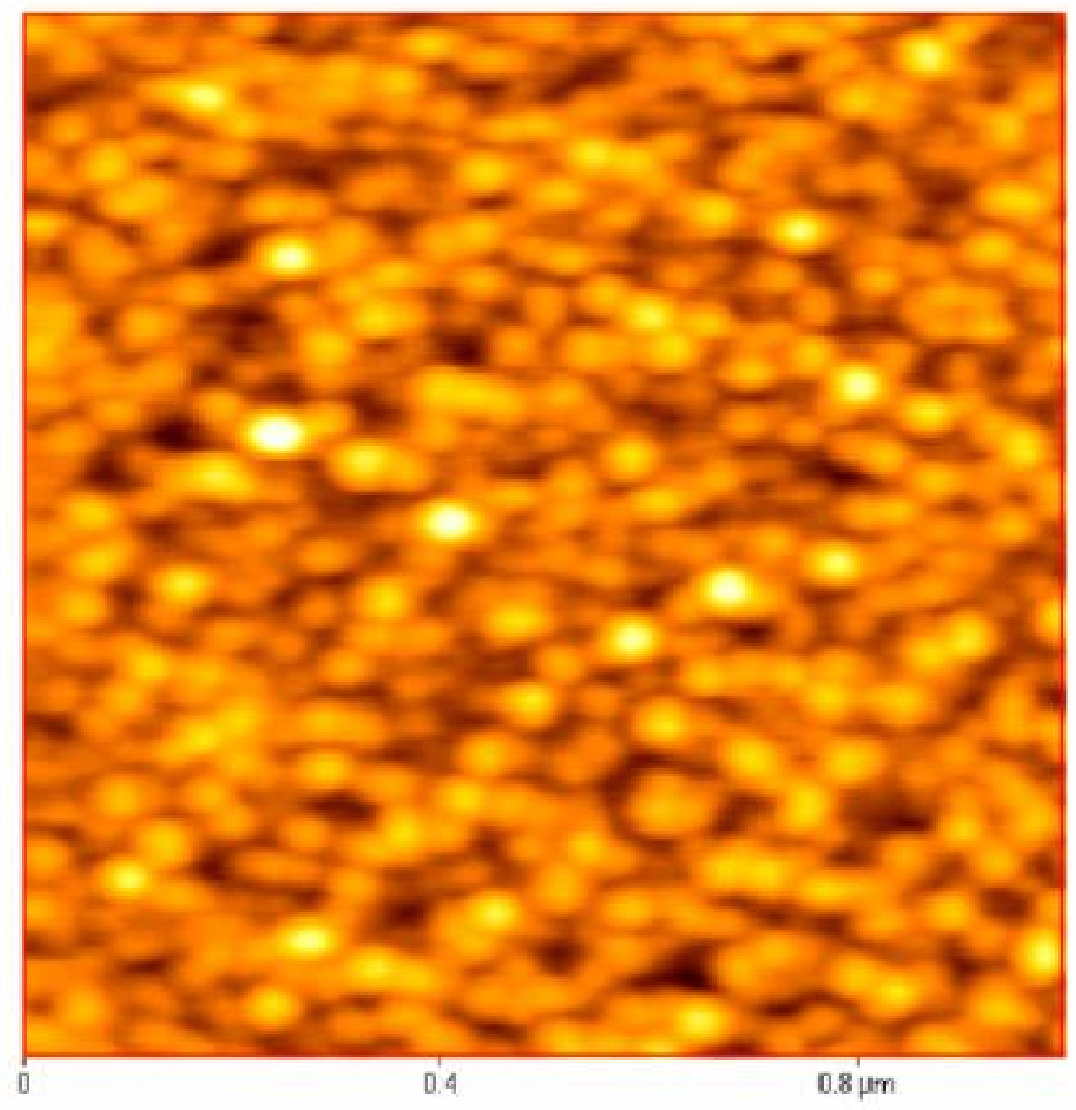}
\includegraphics[width=6cm,height=6cm,angle=0]{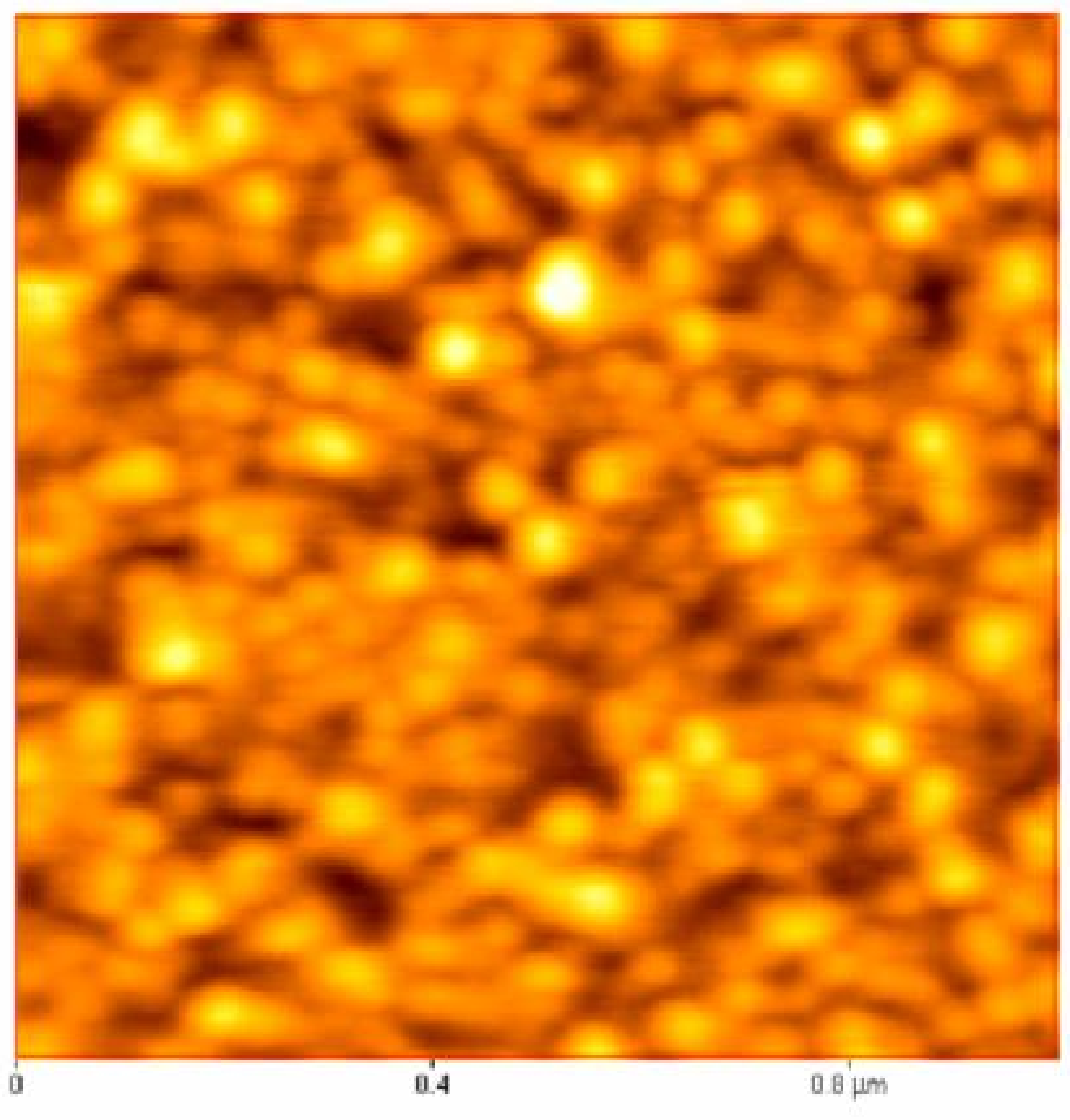}
\includegraphics[width=6cm,height=6cm,angle=0]{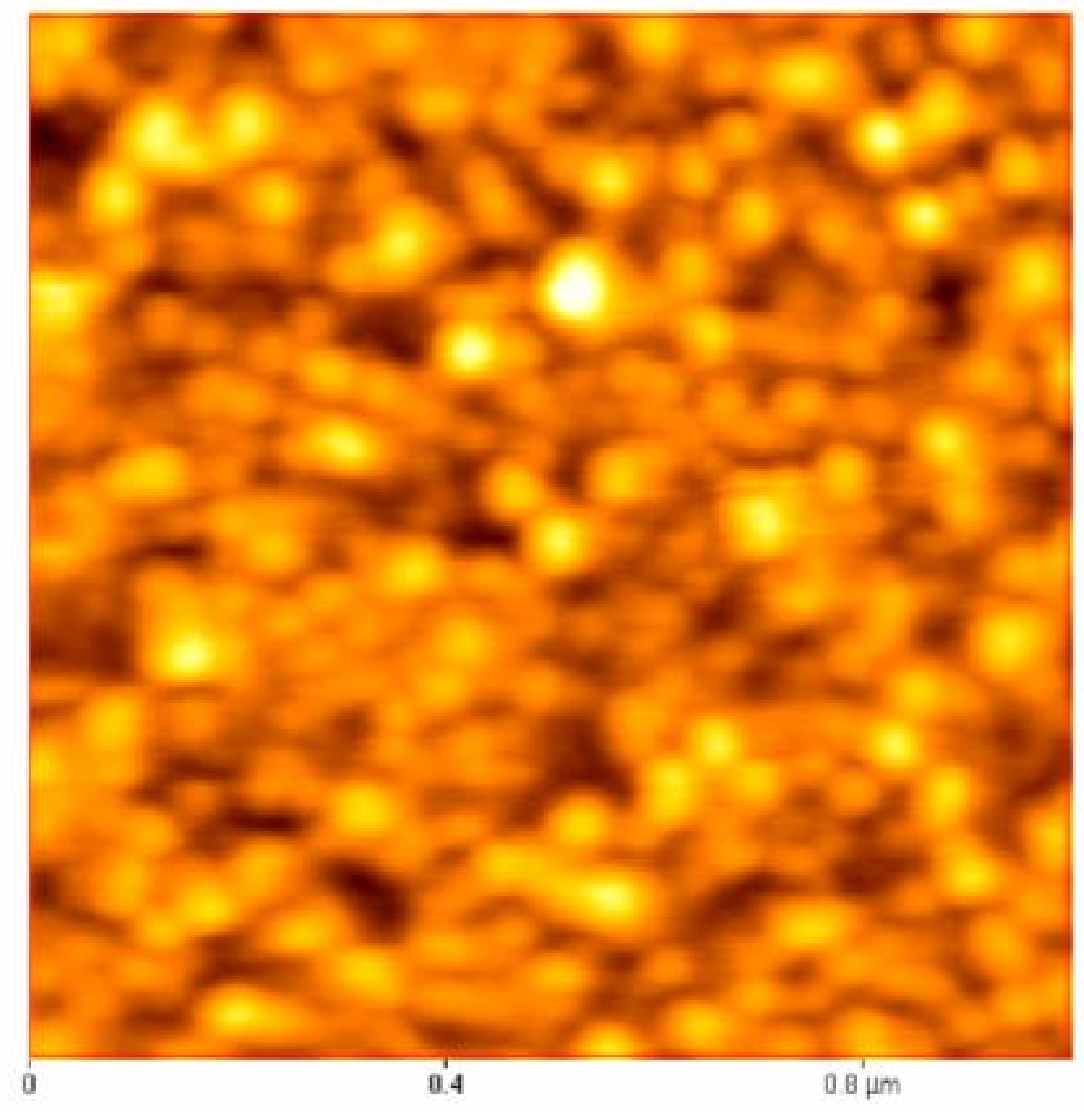}
 \narrowtext \caption{(Color online) AFM surface images  (all
  $ 1\times 1 \mu {\rm m}^2 $) of Co(3
nm)/NiO(30 nm)/Si(100) thin films deposited at the bias voltages
of a) 0, b) -20, c) -40, and d) -60 V. (from top to bottom
corresponding $a$ to $d$, respectively)}
\end{figure}

For the unbiased very thin Co layer, Fig. 1a shows a columnar
structure of the Co grains  grown over the evaporated NiO
underlaying surface. However, Fig. 1b shows that by applying the
negative bias voltage during the Co deposition, the columnar
growth is eliminated. Moreover, Figs. 1c and 1d show that by
increasing the bias voltage up to -60 V the grain size of the Co
layer is increased which means a more uniform and smoother surface
is formed.  But, for the bias voltage of -80 V, due to initiation
of resputtering of the Co surface by the high energy ion
bombardment, we have observed a non-uniform surface, even at the
macroscale of the samples. Therefore, based on the AFM
micrographs, the optimum surface morphology of the Co/NiO/Si(100)
system was achieved at the bias voltage of -60 V for our
experimental conditions \cite{Sangpour}.

Now, by using the introduced statistical parameters in the last
section, it is possible to obtain some quantitative information
about the effect of bias voltage on surface topography of the
Co/NiO/Si(100) system.  Figure 2 presents the structure function
$S(r)$ of the surface grown at the different bias voltages, using
Eq. (\ref{Structure}). The slope of each curve at the small scales
yields the roughness exponent ($\chi$) of the corresponding
surface. Hence, it is seen that the surface grown at the optimum
bias voltage (-60 V) shows a minimum roughness with $\chi=0.60$,
as compared with the other biased samples with $\chi=$0.75, 0.70,
and 0.64 for V=-20, -40, and -80 V, respectively.  For the
unbiased sample, we have obtained two roughness exponent values of
0.73 and 0.36, because of the non-isotropic structure of the
surface (see Fig. 1a).  In any case, at large scales where the
structure function is saturated, Fig. 2 shows the maximum and the
minimum roughness values for the bias voltages of 0 and -60 V,
respectively.

It is also possible to evaluate the grain size dependence to the
applied bias voltage, using the correlation length achieved by the
structure function represented in Fig. 2.  For the unbiased
sample, we have two correlation lengths of 30 and 120 nm due to
the columnar structure of the grains.  However, by applying the
bias voltage, we can attribute just one correlation length to each
curve showing elimination of the columnar structure in the biased
samples.  For the bias voltage of -20 V, the correlation length of
$r^{\ast}$ is found to be 56 nm.  By increasing the value of the
bias voltage to -40 and -60 V, we have measured $r^{\ast}$=76 and
95 nm, respectively.  However, at -80 V, due to initiation of the
destructive effects of the high energy ions on the surface, the
correlation length is reduced to 76 nm.  Now, based on the above
analysis, if we assume that V$_{opt}$=-60 V, then the roughness
exponent and the correlation length can be expressed in terms of
the $\lambda$ as follows, respectively,
\begin{eqnarray}
\chi(\lambda)=0.61+0.16\sin^{2}(2\pi\lambda/3.31+1.07)
\end{eqnarray}
\begin{eqnarray}
r^{\ast}(\lambda)=53.50+40.23\sin^{2}(2\pi\lambda/3.00+2.62) (\rm
nm)
\end{eqnarray}
where for $\lambda=0$ with the columnar structure, we have
considered the average values.

To obtain the stochastic behavior of the surface, we need to
measure the drift coefficient $D^{(1)}(h)$ and diffusion
coefficient $D^{(2)}(h)$ using Eq. (\ref{D(k)}). Figure 3 shows
$D^{(1)}(h)$ for the surfaces at the different bias voltages. It
can be seen that the drift coefficient shows a linear behavior for
$h$ as:
\begin{eqnarray}\label{D1}
D^{(1)} (h,\lambda)=-f^{(1)}(\lambda)h
\end{eqnarray}
where
\begin{eqnarray}
f^{(1)}(\lambda)=[0.55+1.30\sin^{2}(2\pi\lambda/3.50+1.40)]\times
10^{-4}.
\end{eqnarray}
\begin{figure}
\includegraphics[width=8.4cm,height=8cm,angle=0]{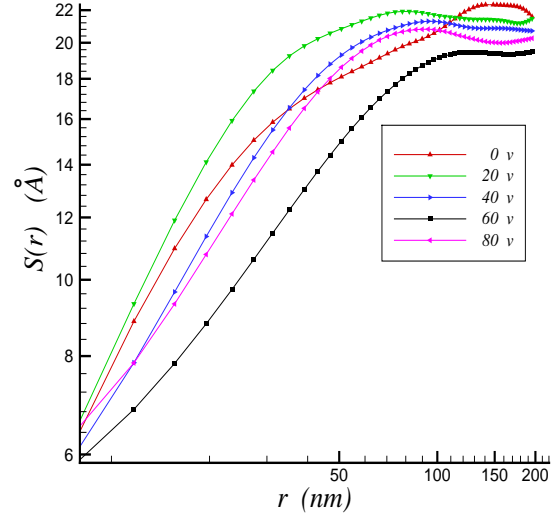}
 \narrowtext \caption{(Color online) Log-Log plot of structure function of the surface at different bias
voltages.}
 \end{figure}
\begin{figure}
\includegraphics[width=8.4cm,height=8cm,angle=0]{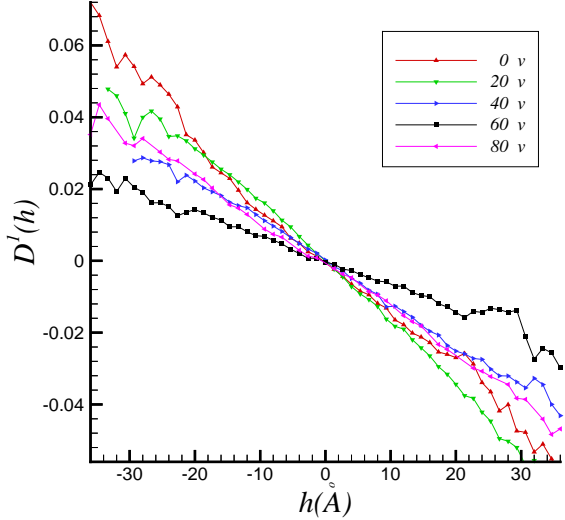}
 \narrowtext \caption{(Color online) Drift coefficient of the surface at different bias
voltages.}
 \end{figure}
The minimum value of $f^{(1)}(\lambda)$ for the biased samples at
$\lambda=1$ shows that the deterministic component of the height
fluctuations for these samples is lower than the other biased and
unbiased ones.
\begin{figure}
\includegraphics[width=8.4cm,height=8cm,angle=0]{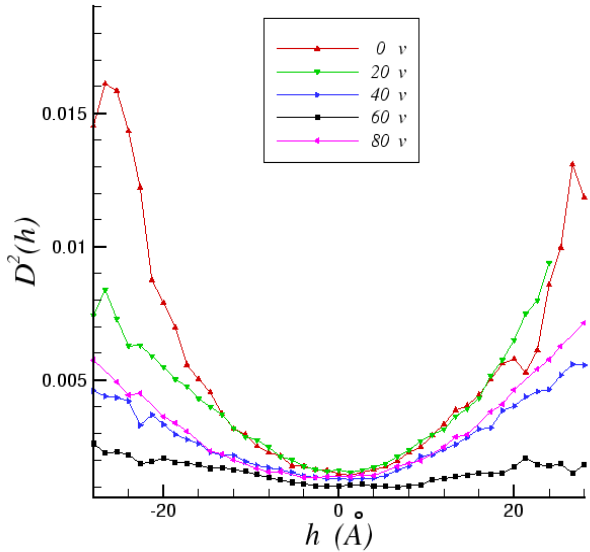}
 \narrowtext \caption{(Color online) Diffusion coefficient of the surface at different bias
voltages.}
 \end{figure}
Figure 4 presents $D^{(2)}(h)$ for the different bias voltages. At
$\lambda=0$, the maximum value of diffusion has been obtained for
any $h$, as compared with the other cases.  By increasing the bias
voltage, the value of $D^{(2)}$ is decreased, as can be seen for
$\lambda=1/3$ and 2/3. The minimum value of $D^{(2)}$, which is
nearly independent of $h$, is achieved when $\lambda=1$.  This shows
that the noisy component of the surface height fluctuation at
$\lambda=1$ is negligible as compared with the unbiased and the
other biased samples.  The behavior of $D^{(2)}$ at $\lambda=4/3$
becomes similar to its behavior at $\lambda=2/3$. It is seen that
the diffusion coefficient $D^{(2)}$ is approximately a quadratic
function of $h$.  Using the data analysis, we have found that
\begin{eqnarray}\label{D2}
D^{(2)} (h,\lambda)=f^{(2)}(\lambda)h^2
\end{eqnarray}
where
\begin{eqnarray}
f^{(2)}(\lambda)=[3.20+3.53\sin^{2}(2\pi\lambda/3.33+1.34)]\times
10^{-6}\nonumber\\
\end{eqnarray}
Now, using the Langevin equation (Eq. (\ref{Langevin})) and the
measured drift and diffusion coefficients, we can conclude that
the height fluctuation has the minimum value at $\lambda=1$ which
means a smoother surface at the optimum condition.  Moreover, the
obtained equations for the coefficients (Eqs. (\ref{D1}) and
(\ref{D2})) can be used to regenerate the rough surfaces the same
as AFM images shown in Fig. 1 \cite{Jafari,Waechter}.

\begin{figure}
\includegraphics[width=8.4cm,height=8cm,angle=0]{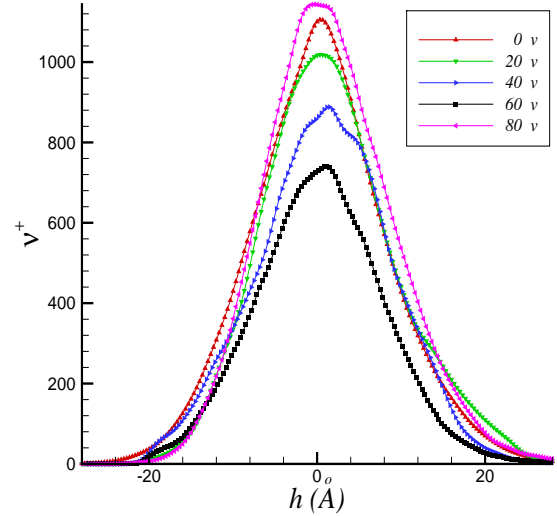}
 \narrowtext \caption{(Color online) Level crossing of the surface at different bias
voltages.}
 \end{figure}

To complete the study, roughness of a surface can be also
evaluated by the level crossing analysis, as another procedure.
Figure 5 shows the observed average frequency $\nu^{+}_{\alpha}$
as a function of $h$ for the different bias voltages.  As
$\lambda$ is increased from 0 to 1, the value of
$\nu^{+}_{\alpha}$ is decreased at any height.  Once again, the
optimum situation is observed for the bias voltage of -60 V
showing the surface formed at $\lambda=1$ condition is a smoother
surface with lower height fluctuations than the surface formed at
the other conditions.  It is seen that, at $\lambda=4/3$, the
height fluctuation of the surface finds a maximum value, as
compared with the other surfaces.  The same as the roughness
exponent and the correlation length behavior in terms of
$\lambda$, the $N^+_{tot}$ can be also expressed as:
\begin{eqnarray}
N^+_{tot}(\lambda)=[1.20+0.17\sin^{2}(2\pi\lambda/3.52+1.40)].
\end{eqnarray}

\begin{figure}
\includegraphics[width=8.4cm,height=8cm,angle=0]{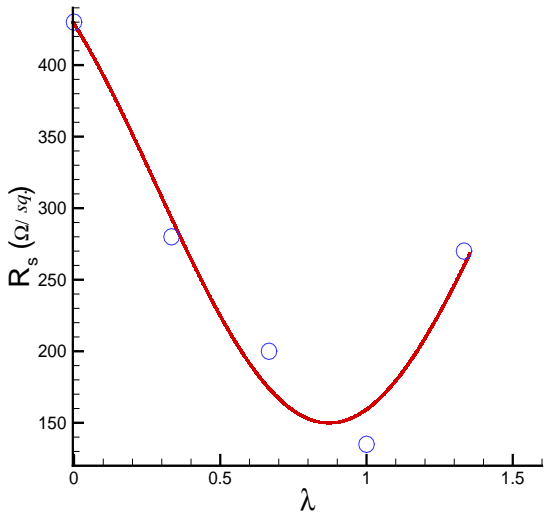}
 \narrowtext \caption{(Color online) Sheet resistance measurement of the Co thin layer as a
function of the applied bias voltage.}
 \end{figure}

Since the system under investigation has a thin Co layer which is
the only conductive layer, thus, it is obvious that lower height
fluctuation corresponds to smaller electrical resistivity of the
surface.  Concerning this, we have measured sheet resistance of the
Co surface grown at the different bias voltages, as shown in Fig. 6.
For the bias voltage ranging from 0 to -60 V, the R$_s$ value is
reduced from 432 to 131 $\Omega / sq.$. The minimum value of R$_s$
is measured at the optimum condition of -60 V $(\lambda=1)$ which
can be related to modified and smooth surface roughness. Elimination
of interfacial voids as well as porosities, and reduction of
impurities in the Co layer. A similar behavior was also observed at
V$_{opt}$=-50 V for Ta/Si(111) system \cite{AkhavanTSF}. By
increasing the applied bias voltage to values greater than its
optimum value, surface roughness is increased because of surface
bombardment by high energy ions. This can be seen by the observed
increase in the R$_s$ value at the bias voltage of -80 V
($\lambda=4/3$).  It is easy to examine that the variation of R$_s$
as a function of $\lambda$ can be expressed as below:
\begin{eqnarray}
R_s(\lambda)=[135.48+307.74\sin^{2}(2\pi\lambda/3.93+1.77)]
\end{eqnarray}
It behaves similar to the behavior of roughness characteristics of
the surfaces. Therefore, we have shown that the roughness behavior
explained by the statistical characterizations of the surface,
which have been obtained by using microstructural analysis of AFM,
can be related to the sheet resistance measurement of rough
surfaces, as a macrostructural analysis.

\section{The tip convolution effect}

It is well-known that images acquired with AFM are a convolution
of tip and sample interaction.  In fact, using scanning probe
techniques for determining scaling parameters of a surface leads
to an underestimate of the actual scaling dimension, due to the
dilation of tip and surface. Concerning this, Aue and Hosson
\cite{Aue} showed that the underestimation of the scaling exponent
depends on the shape and aspect ratio of the tip, the actual
fractal dimension of the surface, and its lateral–vertical ratio.
In general, they proved that the aspect ratio of the tip is the
limiting factor in the imaging process.

Here, we want to study the aspect ratio effect of the tip on the
investigated stochastic parameters. To do this, using a computer
simulation program, we have generated a rough surface by using a
Brownian motion type algorithm \cite{Peitgen,24} with roughness
and its exponent of 10.00 nm and 0.67, respectively. We have
assumed these roughness parameters in order to have some
similarity between the generated surface and our analyzed surface
by AFM. In the simulation program, the generated surface has been
scanned using a sharp cone tip with an assumed aspect ratio of
$0.73$ which is also nearly similar to the applied tip in our AFM
analysis with the aspect ratio of 0.9. Moreover, this assumption
does not limit the generality of our discussion, because it is
shown that the fractal behavior of a rough surface presents an
independent tip aspect ratio behavior (saturated behavior) for the
aspect ratios greater than about 0.4 \cite{Aue}.
\begin{figure}
\includegraphics[width=8.4cm,height=8cm,angle=0]{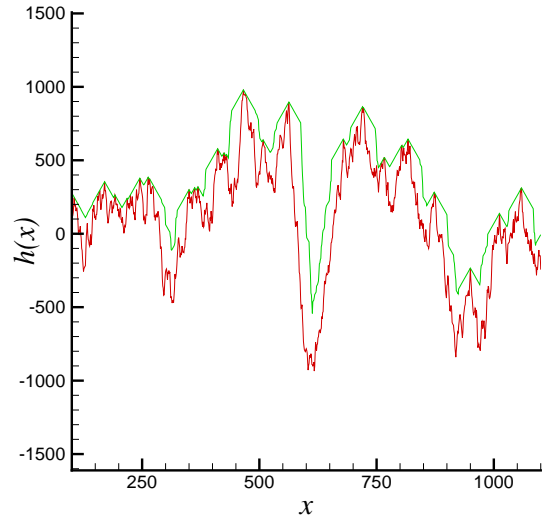}
 \narrowtext \caption{(Color online) Height profile of a rough generated surface before
dilation (real) and after dilation (tip) caused by a tip with the
aspect ratio of 0.73.}
 \end{figure}

Figure 7 shows a line profile of a generated surface which is
dilated by a tip with the known aspect ratio.  It is clearly seen
that the scanned image (the image affected by the tip convolution)
does not completely show the generated surface topography (real
surface). Now, it is possible to study the dependance of the
examined surface stochastic parameters on the geometrical
characteristic of the tip, i.e. aspect ratio.

\begin{figure}
\includegraphics[width=8.4cm,height=8cm,angle=0]{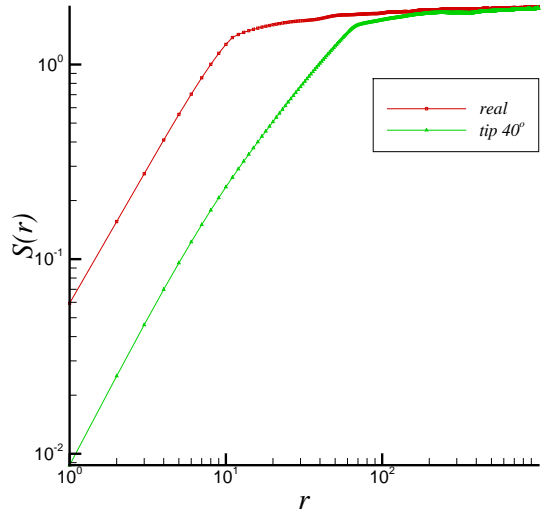}
 \narrowtext \caption{(Color online) The one-dimensional structure function analysis,
plotting log[$S(r)$] $vs.$ log($r$) in which $r$ is pixel position
along the $x$-axis.  This results in the roughness values of 10 and
7.36 nm for the generated surface before dilation and after dilation
using a tip with the aspect ratios of 0.73, respectively.}
 \end{figure}

In this regard, Fig. 8 shows variation of the one-dimensional
structure function of the generated rough surface due to the tip
convolution effect. It can be seen that by increasing the aspect
ratio the tip convolution results in obtaining a surface image
whit a decreased roughness. Since the aspect ratio of the applied
AFM tip was around 0.9, so the measured roughness exponents at the
different bias voltages might be corrected by a 1.07 factor. In
other words, the relative change (the difference between the real
and measured values comparing the real one) of the roughness
exponent is about 7.2\%.
\begin{figure}
\includegraphics[width=8.4cm,height=8cm,angle=0]{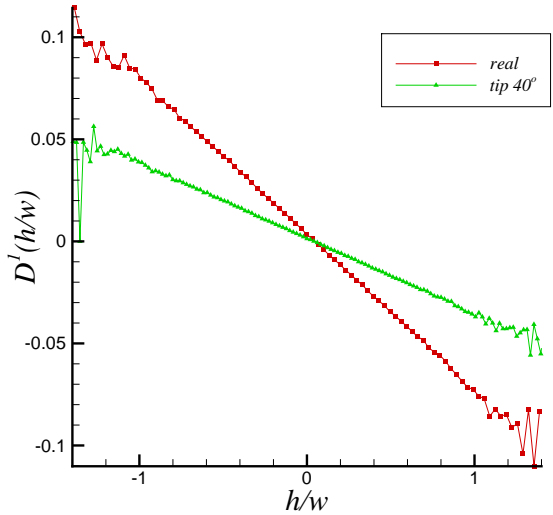}
 \narrowtext \caption{(Color online) The calculated drift coefficient for the generated
surface before dilation (real) and after dilation with a tip having
aspect ratio of 0.73 (tip).}
 \end{figure}
Moreover, Fig. 8 shows that the correlation length is increased by
the tip convolution effect.  It should be noted that, in our
simulation, we have assumed that the apex of the tip is completely
sharp (the tip radius is assumed zero).  However, it is well-known
that the radius of the pyramidal tips is $\sim$ 20 nm. Therefore,
the real correlation lengths are even roughly 20 nm larger than
the measured ones by the sharp tip.

\begin{figure}
\includegraphics[width=8.4cm,height=8cm,angle=0]{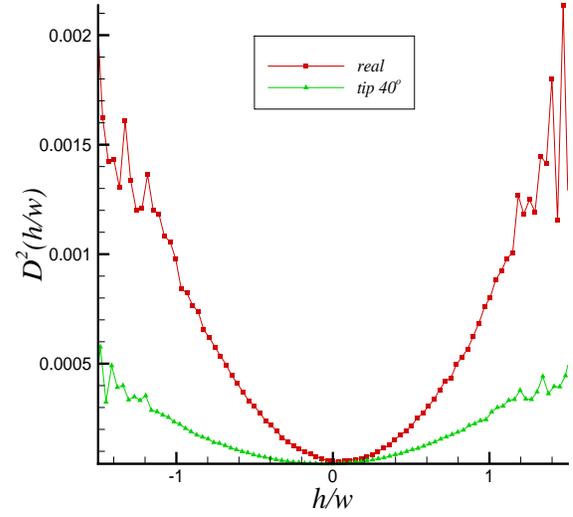}
 \narrowtext \caption{(Color online) The calculated diffusion coefficient for the generated
surface before dilation (real) and after dilation with a tip having
aspect ratio of 0.73 (tip).}
 \end{figure}
\begin{figure}
\includegraphics[width=8.4cm,height=8cm,angle=0]{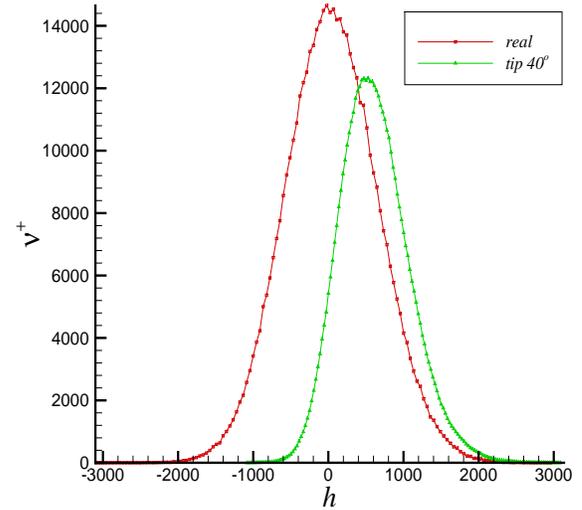}
 \narrowtext \caption{(Color online) Level crossing analysis of the generated surface before
dilation (real) and after dilation (tip).}
 \end{figure}

The same tip convolution effect can be also presented for the
drift and diffusion coefficients. Figure 9 presents the calculated
drift coefficient for the generated surface and the scanned
surface. One can see that the tip convolution results in
decreasing of the drift coefficient, corresponding to decreasing
of the surface roughness. This means that after dilation the
correlation length will increase and hence the measured value for
$f^{(1)}(\lambda)$ would be smaller than its value for the
original surface. Therefore, the magnitude of slope of the drift
coefficient must decrease after using the tip. For our generated
surface, the measured value of the drift coefficient should be
modified by a factor of around 2.

The variation of the diffusion coefficient of the generated rough
surface due to the tip convolution effect has been also shown in
Fig. 10.  The reduction of the diffusion coefficient of the
scanned surface as compared to its values for the generated
surface, due to the tip convolution, can be easily seen.  In fact
to compensate the tip effect on the diffusion coefficient, we
should modify its measured values by a factor of about 4, for the
assumed generated surface.

Finally, we remind that the total number of crossing the surface
with positive slope ($N^{+}_{tot}$) has been defined as a
parameter describing the rough surfaces.  Hence, we have also
studied the effect of the tip convolution on this parameter, as
shown in Fig. 11.  It is seen that $N^{+}_{tot}$ decreased due to
the tip convolution effect.  For the assumed generated surface, we
have obtained that the $N^{+}_{tot}$ of the surface before
dilation is about 1.7 times larger than its value after the
dilation.  In this figure, we have also shown the variation of the
average height due to the tip effect.

One has to note that our generated surface is a pure
two-dimensional one which presents no line-to-line interaction.
So, for this simple model, differently shaped tips with the same
aspect ratio yield the same results. Therefore, for the
three-dimensional case one can expect to obtain a larger
distortion of the surface due to stronger line-to-line interaction
leading to an even larger underestimation of the studied
stochastic parameters.

These analysis showed that, although the measured values of the
surface parameters by AFM method are different from the real ones,
the general behavior of these parameters as a function of the bias
voltage are not affected by the tip convolution.   Therefore, our
general conclusions about the variation of the studied stochastic
parameters by applying the bias voltage is intact.

\section{Conclusions}

We have investigated the role of bias voltage, as an external
parameter, to control the statistical properties of a rough surface.
It is shown that at an optimum bias voltage ($\lambda=1$), the
stochastic parameters describing a rough surface such as roughness
exponent, level crossing, drift and diffusion coefficient must be
found in their minimum values as compared to an unbiased and the
other biased samples.  In fact, dependence of the height fluctuation
of a rough surface to different kinds of the external control
parameters, such as bias voltage, temperature, pressure, and so on,
can be expressed by AFM data which are analyzed using the surface
stochastic parameters. In addition, this characterization enable us
to regenerate the rough surfaces grown at the different controlled
conditions, with the same statistical properties in the considered
scales, which can be useful in computer simulation of physical
phenomena at surfaces and interfaces of, especially, very thin
layers.  It is also shown that these statistical and microstructural
parameters can explain well the macroscopic properties of a surface,
such as sheet resistance.  Moreover, we have shown that the
tip-sample interaction does not change the physical behavior of the
stochastic parameters affected by the bias voltage.
\begin{acknowledgments}
AZM  would like to thank Research Council of Sharif University of
Technology for financial support of this work. We also thank F.
Ghasemi for useful discussions.
\end{acknowledgments}


\end{document}